# A Control Strategy for Capacity Allocation of Hybrid Energy Storage System Based on Hierarchical Processing of Demand Power

Kai Lin[1]

[1] University of Chinese Academy of Sciences, Yanqihu Campus, Huaibei Zhuang, Huairou District, Beijing,100049, PR China

**Abstract:** Pursuing optimal power distribution in hybrid energy storage systems has always been the goal of researchers. Here, HESS is a combination of lithium battery and supercapacitor; this combination has been proven to effectively compensate for some of the deficiencies of lithium batteries as an energy system for electric vehicles. For example, the energy storage system with only lithium batteries cannot provide high power in a short time to meet the high acceleration performance of electric vehicles, and the excessive discharge current will cause the temperature of the battery pack to be too high, which will cause safety problems for the car. This paper proposes an intelligent energy management strategy combining fuzzy controller and improved Savitzky-Golay filter for real-time control. The simulation results show that compared with single ESS, the maximum current of the battery proposed by the strategy is reduced by 14.60%, and the usable cycle life of the battery is increased by 57.31% during the test driving cycle. Meanwhile, it explores various changes brought supercapacitor monomers in the same HESS, and predict the next supercapacitor will bring about 31.58% reduction of volume and mass.

**Keywords:** Hybrid energy storage system; Fuzzy logic control.

## Ⅰ. Introduction

Further development of electric vehicles (EV) require lithium-ion battery technologies improvement continuously[1] [2] [3] [4]. One of the important concerns is that high power demand with high stable and long life working time is hard to be satisfied together by current LIB technologies [5, 6]. To couple lithium-ion battery with supercapacitors together is recognized as the most popular solution with efficient and reasonable power distribution [7] .Depending on the features of ultra-high power and ultra-long working life time, supercapacitors can be used to cut and provide peak power, which can significantly reduce the requirements of more expensive power type lithium-ion battery and the damage of high current to the working life and stability of lithium-ion battery, and higher energy efficiency also can be achieved [8, 9] [10] [11]. In view of the nonlinear characteristics of electric vehicle power demand, this work develop an intelligent real-time energy management strategy through fuzzy logic control with an enhanced Savitzky-Golay filter. Through the analysis of the power demand of the passenger car, the proposed control strategy can effectively reduce the loss of the hybrid energy storage system and improve the economy of the energy storage system.

In order to enhance the performance of HESS, various power control strategies have been proposed, and generally could be classified as rule-based and optimisation-based strategies [12, 13] [14] [15]. While the another fuzzy-based control strategy is developed based on the engineer experiences[16] [17]. A simple table lookup method is introduced [18]. The method is divided into two modes, each used to provide power in the form of different power requirements [19]. Bowman et al. [20] and Lee and Sul [21] proposed a torque control strategy for parallel electric vehicles based on fuzzy rules as early as 1998. Zhang et al.[10]

proposed a filtering method to provide the high-frequency part of the required power to the supercapacitor and the low-frequency part to the lithium-ion battery. Compared with the traditional control strategy, it effectively increases the life of the energy storage system and reduces the power loss of the energy storage system.

The degradation and system loss cost of the HESS are usually regarded as the evaluation function of offline optimization, and optimal power distribution point is determined depending on priori data of entire drive cycles such as the non-dominated sorting genetic algorithm-II(NSGA-II) [22], dynamic programming (DP) [23, 24]. Because these non-real time solutions rely on initial drive data and more computing resources, they are merely serve as benchmarks for achieving optimal system performance. While, some global optimization methods, such as genetic algorithm[25, 26], particle swarm optimization [27], simulated annealing [6][28], can be used for the development of optimal controllers.

Four topological ways are popular used to couple the lithium battery and supercapacitor in HESS, a passive, two semi-active and fully active topological structures. Although semi-active topology (Fig.1) only uses one DC / DC converter, it can provide the better balance between costs and performances, and thus it was used here.

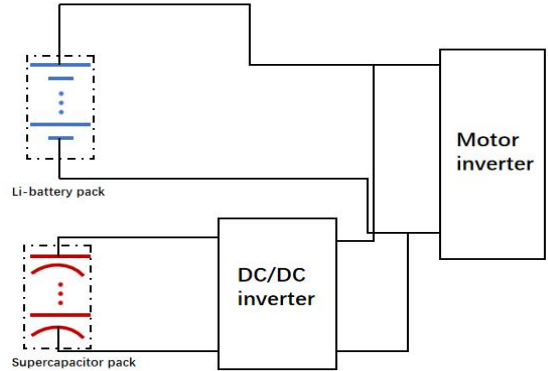

**Fig.1.** Supercapacitor controlled semi-active topological hybrid system.

## II. HESS MODELING

A. Dynamic Model of the Battery

The battery model used in this paper is the PNGV equivalent circuit model proposed by the United States' new-generation automotive cooperation plan in 2001. The PNGV model adds a capacitor $C_p$ based on the Thevenin model to describe the open-circuit voltage change generated by the cumulative load current time. As shown in Fig.2. Where $U_{oc}$ represents an ideal voltage source, and is used to describe the battery open-voltage; $R_p$ is the ohm resistance; Polarisation $R_s$ and capacitance $C_s$ describe the battery over-voltage $U_p$; $U_L$ and $I_{bat}$ are the load voltage and current of the battery, respectively. The basic parameters of the tested battery pack are listed in Table I.

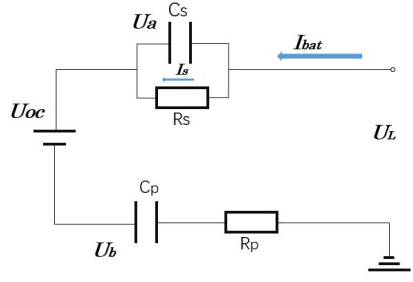

**Fig.2.** PNGV equivalent circuit model of the battery.

TABLE I
BASIC PARAMETERS OF THE BATTERY

| Parameters | Value | Unit |
|---|---|---|
| Battery Type | Lithium-ion battery | |
| Nominal pack voltage | 330 | V |
| Nominal pack capacity | 40 | Ah |
| Parallel number | 2 | |
| Normal cell capacity | 20 | Ah |

Based on the equivalent circuit, the following voltage state equation is established:

$$U_{oc} = U_L - R_S I_S - \frac{1}{C_P}\left(\int I_{bat}\, dt\right) - I_{bat} R_p \quad (1)$$

$$\frac{dI_s}{dt} = \frac{I_L - I_S}{\tau} \quad (2)$$

$$\tau = R_S C_S \quad (3)$$

Where $\tau$ denotes the time constant. The remaining charge SOC(t) of a battery at a specific time (t) is usually calculated based on the initial time $SOC_0$. By calculating the integral of current to time in time (t), the percentage of remaining power can be obtained. which is defined by

$$SOC(t) = SOC_0 \frac{1}{Q_N}\int_0^t \eta\, I(t)\, dt \quad (4)$$

Among them, $Q_N$ is the battery rated capacity, $I(t)$ is the battery current in time (t), and $\eta$ is the charge and discharge efficiency.

The calculation method of the total actual power loss of the battery pack during time $\lambda$ is as follows:

$$Q_{loss} = \sum_{t=0}^{\lambda} (I(t)^2 R/m - U(t)I(t)(1-Q_c)) \quad (5)$$

Where $I(t)$ is the output current of the battery pack at time t, R is the resistance of the battery pack, $U(t)$ and $\eta$ are the terminal voltage and Coulomb efficiency of the battery pack (coulomb efficiency of ultracapacitor $\eta_{sc}$ =0.95; coulomb efficiency of lithium battery $\eta_{bat}$=0.85.). The m is the number of parallel branches of the battery pack.

B. Dynamic Model of the Supercapacitor

In order to describe the dynamic characteristics of the supercapacitor, the RC equivalent circuit model is used to model the supercapacitor, as shown in Fig. 3. The Model includes a large capacitor $C_b$ and a characteristic small capacitor $C_s$. The $C_b$ and $C_s$ are used to simulate the static and dynamic states of the supercapacitor. $R_t$, $R_s$ and $R_e$ are surface resistances respectively, and are used to reflect the storage capacity and dynamic characteristics of the supercapacitor.

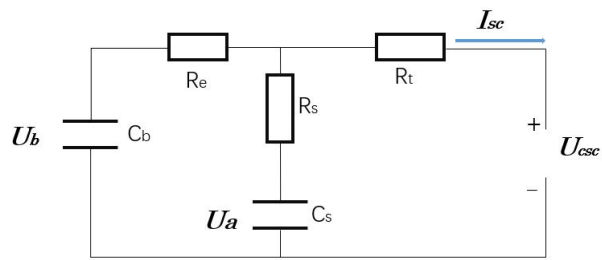

**Fig.3.** RC equivalent circuit model of the supercapacitor.

According to the above RC battery equivalent circuit, the mathematical model is compiled as follows.

$$\begin{bmatrix} \dfrac{dU_b}{dt} \\ \dfrac{dU_s}{dt} \end{bmatrix} = \begin{bmatrix} \dfrac{-1}{C_b(R_e + R_s)} \\ \dfrac{1}{C_s(R_e + R_s)} \end{bmatrix} \cdot \begin{bmatrix} U_b \\ U_s \end{bmatrix} + \begin{bmatrix} \dfrac{-R_s}{C_b(R_e + R_s)} \\ \dfrac{-R_e}{C_s(R_e + R_s)} \end{bmatrix} \cdot [I_{sc}] \quad (6)$$

$$U_{sc} = \frac{R_s - R_t}{R_e + R_s} U_b + \frac{R_e + R_t}{R_e + R_s} U_s + \frac{R_e R_t - R_e R_s}{R_e + R_s} I_{sc} \quad (7)$$

The basic parameters of the supercapacitor are listed in Table II.

TABLE II
BASIC PARAMETERS OF THE SUPERCAPACITOR

| Parameters | Value | Unit |
| --- | --- | --- |
| Battery Type | Supercapacitor | |
| Nominal pack voltage | 240 | V |
| Nominal pack capacity | 34 | F |
| Parallel number | 1 | |
| Normal cell capacity | 3000 | F |

The $SOC$ of the supercapacitor at the current moment can be expressed as

$$SOC = \frac{Q_{remaining}}{Q_{total}} = \frac{C * (U_{sc} - U_{min})}{C * (U_{max} - U_{min})} = \frac{U_{sc} - U_{min}}{U_{max} - U_{min}} \quad (8)$$

Where $Q_{remaining}$ is the amount of charge remaining in the current supercapacitor, $Q_{total}$ is the total capacitor when the supercapacitor bank is full charge. $U_{min}$ and $U_{max}$ represent the minimum and maximum voltages across the capacitor, respectively.

### III. Energy management strategy

The fuzzy controller has strong robustness and is suitable for solving the problems of non-linearity, strong coupling, time-varying, and hysteresis in process control. It is widespread in solving real-time control problems. Here, a control strategy is proposed that combines a fuzzy controller with an improved Savitzky-Golay filter. The overall flow chart and a specific block diagram of this control strategy's energy management system are shown in Fig.4.

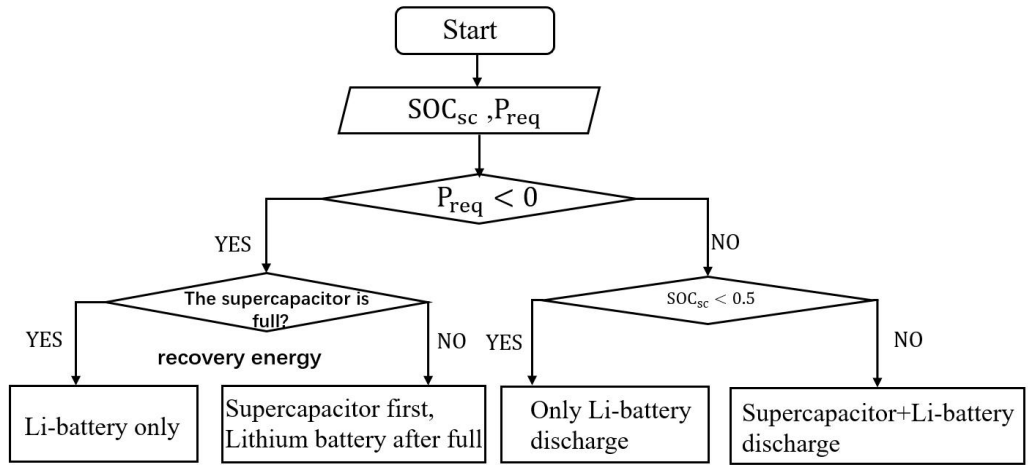

(a). Flowchart for an energy management system.

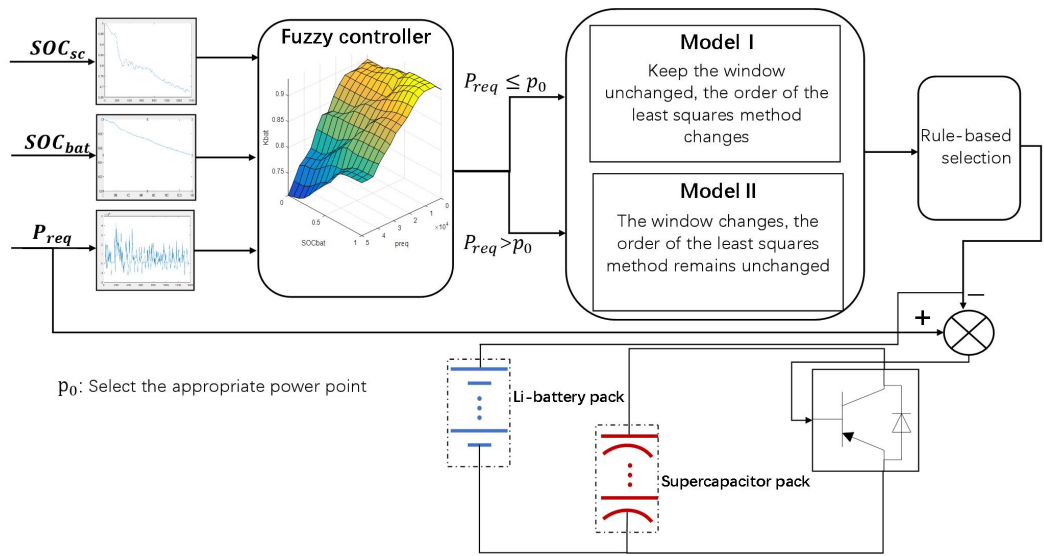

(b). Diagram of the proposed energy management system.

**Figure.4.** Flow chart of energy management of hybrid energy storage system.

A. Design of Fuzzy Controller

The proposed fuzzy controller requires three inputs, the required power $P_{req}$, the $SOC$ of battery and supercapacitor with on output, $K_{bat}$, that is the power ratio of the provided by the lithium battery to the total. In the developed fuzzy logic controller, the relationship between the input and output of the fuzzy controller is shown in Fig.5 (a) and (b).

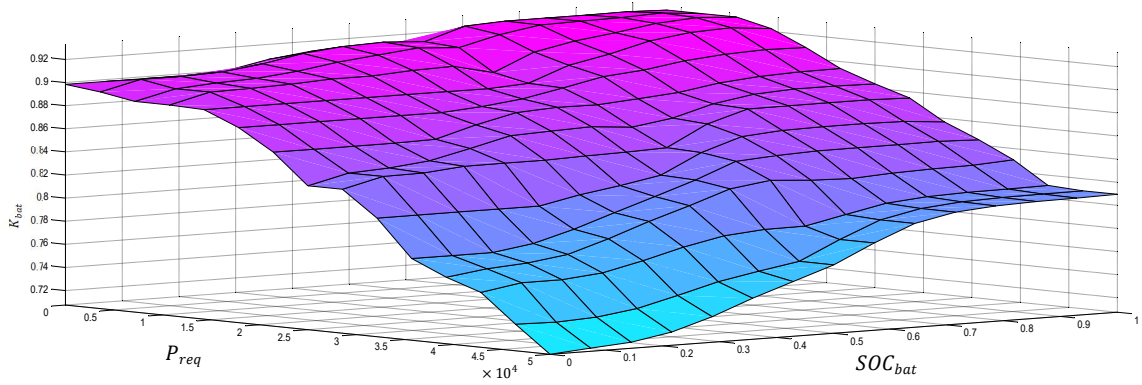

(a) The relationship between the input and output of the fuzzy controller

$(SOC_{bat}, P_{req}$ and $K_{bat})$.

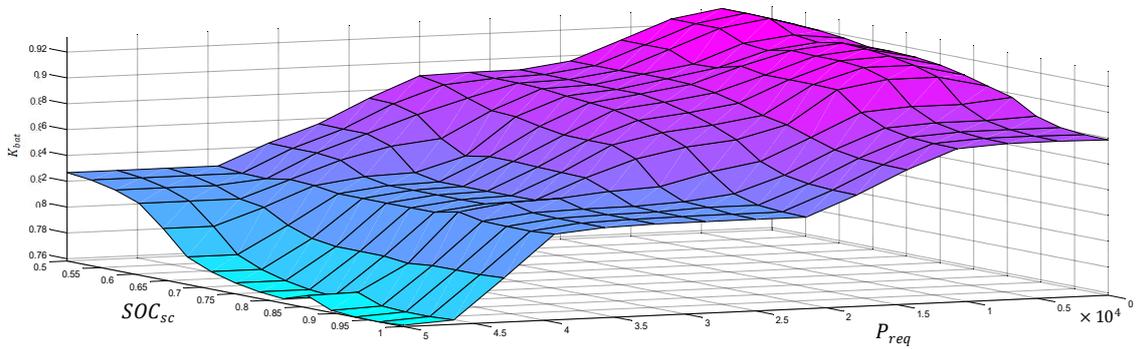

(a) The relationship between the input and output of the fuzzy controller

$(SOC_{sc}, P_{req}$ and $K_{bat})$.

**Fig.5.** The relationship between the input and output of the fuzzy controller.

B. Improved Savitzky-Golay filter

Savitzky-Golay filter is widely used in data stream smoothing and noise removal [29]. It's most prominent feature is that it can ensure the shape and width of the signal unchanged while filtering the noise. The steps to smooth the output power of renewable energy using the Savitzky-Golay filter are as follows.

The width of the filter window is set as n=2m+1, and each measurement point is x=(-m,-m+1,······,0,1,······m-1,m), and using (k-1) order polynomial to fit the data points in the window

$$y(x) = a_0 + a_1 x + a_2 x^2 + a_3 x^3 + \cdots + a_{k-1} x^{k-1} = \sum_{t=1}^{k} a_{t-1} x^{t-1} \quad (9)$$

Thus there are n such equations with forming k elements linear equation system. To make the equations possess the solution, n should be greater than or equal to k. Generally, n>k is selected, and the fitting parameter A is determined by the least-squares method. It can be

obtained

$$\begin{pmatrix} y(-m) \\ y(-m+1) \\ \vdots \\ y(m) \end{pmatrix} = \begin{pmatrix} 1 & -m & \cdots & (-m)^{k-1} \\ 1 & -m+1 & \cdots & (-m+1)^{k-1} \\ \vdots & \vdots & \vdots & \vdots \\ 1 & m & \cdots & (m)^{k-1} \end{pmatrix} \begin{pmatrix} a_0 \\ a_1 \\ \vdots \\ a_{k-1} \end{pmatrix} \quad (10)$$

The matrix can be represented as follows

$$Y_{(2m+1)\times 1} = X_{(2m+1)\times k} \cdot A_{k\times 1} \quad (11)$$

The **E** matrix is

$$E = X_{(2m+1)\times k} \quad (12)$$

Set the other auxiliary matrix $B$ as

$$B = E^T \cdot E \quad (13)$$

The least square solution A of $A_{k\times 1}$ is

$$A = (E^T \cdot E)^{-1} \cdot E^T \cdot Y \quad (14)$$

The predicted value $\breve{Y}$ of $Y$ is

$$\breve{Y} = E \cdot A \quad (15)$$

The goodness of fit (denoted as $R^2$) as an output evaluation can be expressed as

$$R^2 = 1 - \frac{SS_{res}}{SS_{tot}} \quad (16)$$

Assuming a data set includes n recorded values of $k_0, k_1, \cdots k_n$. And the corresponding model prediction values are $h_0, h_1, \cdots h_{n-1}$ and $h_n$ respectively.

$$\bar{y} = \frac{1}{n+1} \sum_{i=0}^{i=n} k_i \quad (17)$$

Then the total sum of squares can be obtained as

$$SS_{tot} = \sum_{i=0}^{i=n} (k_i - \bar{y})^2 \quad (18)$$

The regression sum of squares is

$$SS_{res} = \sum_{i=0}^{i=n} (k_i - h_i)^2 \quad (19)$$

Therefore, the goodness of fit can be defined as

$$R^2 = 1 - \frac{SS_{res}}{SS_{tot}} \qquad (20)$$

In the case of Mode I, the window is fixed (the collected data is constant), and the order of the polynomial (9) changes. If there are c windows, the order of the polynomial varies within [1,t-1]. As shown in Fig.6.

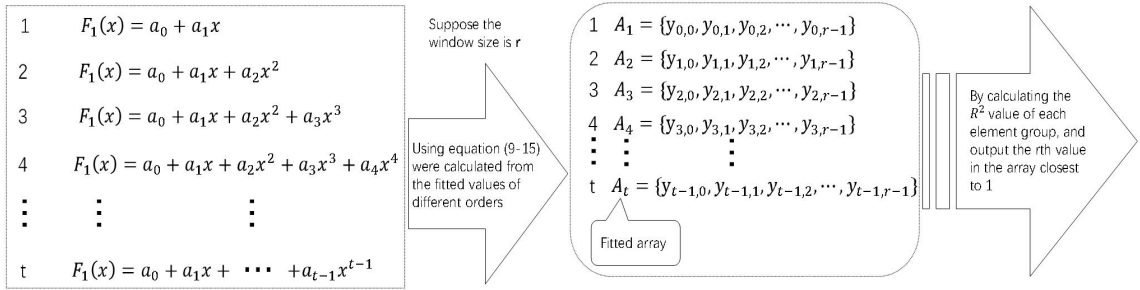

**Fig.6.** Schematic diagram of the calculation process of Model I

In mode II, the window is changing (the number of collected data is different), and the order of polynomial (9) remains unchanged. As shown in Fig.7.

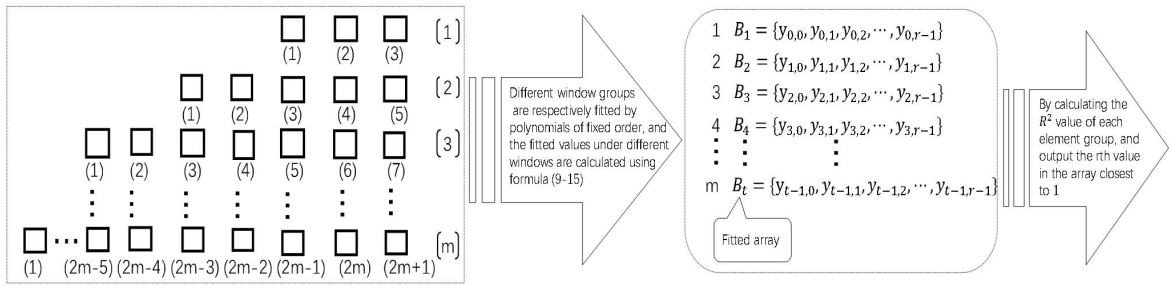

**Fig.7.** Schematic diagram of the calculation process of Model II

Rule-based selection is to prevent Model I and model II from overfitting through simple rules. When the output value of $P_{bat}$ is larger than the required power, which output may cause excessive power loss of the lithium battery, so at this time, the value of the fuzzy controller is output by the design of relevant rules.

## IV. Simulation Results and Discussions

The simulation was implemented in Matlab environment for verifying the effectiveness of the strategy. The urban dynamometer driving schedule (CYC_UDDS) was selected as the test driving cycle; the specific characteristics are shown in Fig.8, and the main parameters of the vehicle are listed in Table III.

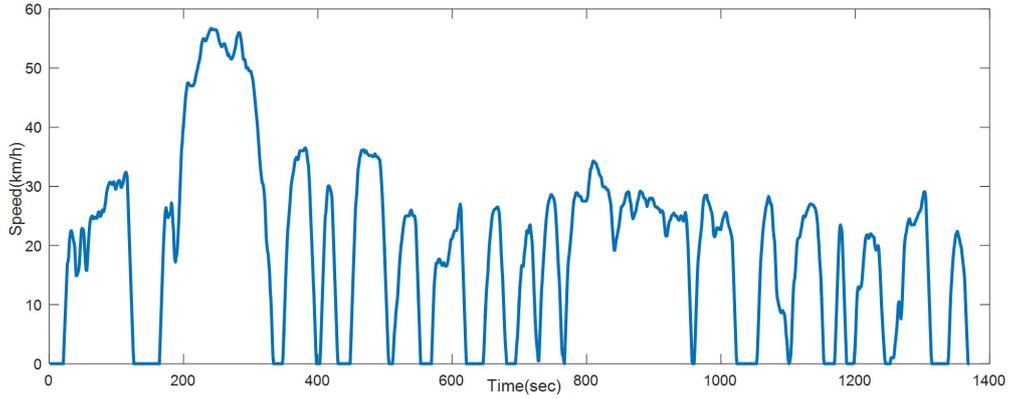

**Fig.8.** test driving cycle

TABLE III

MAIN PARAMETERS OF THE VEHICLE

| Parameters | Value |
|---|---|
| Mass(kg) | 1635 |
| Frontal area($m^2$) | 2.04 |
| Tire radius(m) | 0.28 |
| Drag coefficient | 0.41 |
| Rolling resistance coefficient | 0.03 |

The energy loss ($Q_{loss}$) of HESS is an essential indicator of the coordinated working efficiency of the battery and supercapacitor HESS. When the battery and supercapacitor are charged and discharged within the allowable capacity, the energy loss of the energy storage system is small (under agreed cycling conditions), indicating that the strategies can effectively allocate the power. As shown in table IV, the energy loss of HESS with the proposed strategy was reduced by 27.53% in comparing with the fuzzy strategy.

TABLE IV

ENERGY LOSS OF ENERGY STORAGE SYSTEM

| Strategies | $Q_{bat-loss}(KJ)$ | $Q_{sc-loss}(KJ)$ | $Q_{loss}(KJ)$ |
|---|---|---|---|
| Single ESS | 1395.95 | 0.00 | 1395.95 |
| Proposed Strategy | 983.91 | 27.80 | 1011.71 |

The following figures show the changes in various performance indicators of batteries and supercapacitors under the two strategies.

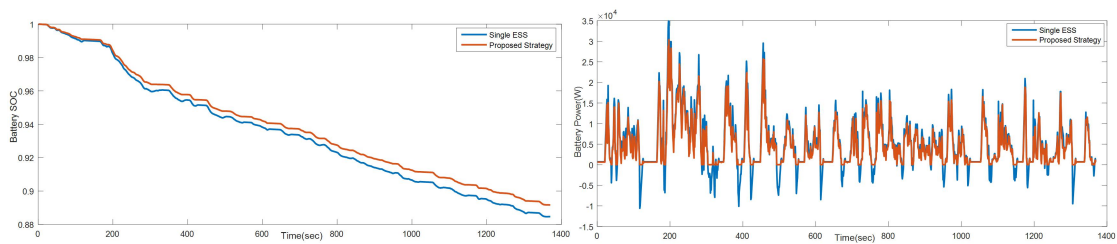

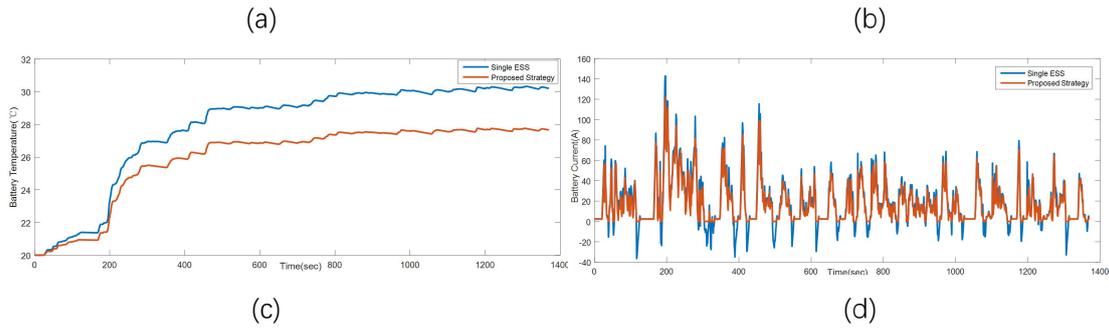

(a)                                                                (b)

(c)                                                                (d)

**Fig.9.** Comparison of various performances of batteries in energy storage systems under two strategies

As can be seen in Fig. 9 (a), under the proposed strategy, the SOC of the battery in the hybrid energy storage system is higher than that under single ESS; their values are 0.8852 and 0.8949 respectively, this is mainly because more peak power will be allocated to the supercapacitor after the fuzzy controller output power is fitted by improved Savitzky-Golay filter. The output power of the strategy proposed in Figure (b) is smaller than that of the single ESS, especially at peak power. Figure (c) shows the temperature change curve of the battery. The maximum temperature is reduced 8.47% in comparing with that of single ESS, and the final temperature drops by 8.41% after the test cycles (Fig. 9c), this decrease helps to improve the safety of the battery pack. The highest current of proposed method is 122.06A and reduced by 14.60% (Fig. 9d). The decrease of battery output current is beneficial to slow down battery degradation and improve service life.

    The fluctuation of battery output current is closely related to battery degradation[30] [2]. Excessive current fluctuations between neighbors will accelerate battery degradation and reduce battery service life. Then the absolute difference $\delta I$ between adjacent currents is as follows

$$\delta I = |I(t) - I(t-1)| \quad\quad (21)$$

Where $I(t)$ is the current value at time t, and $I(t-1)$ is the instantaneous current at the time (t-1). In the test cycle, the calculated $\delta I$ distribution image is shown in Fig.10.

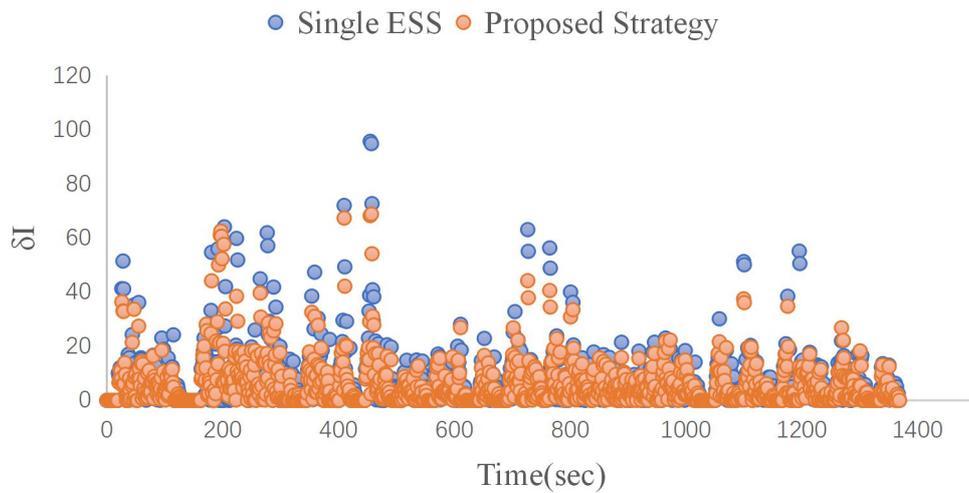

**Fig.10.** Distribution of battery output current difference $\delta I$.

It can be seen from Fig 10 that the ordinate of the current difference output between adjacent batteries in the proposed strategy is smaller than that of the fuzzy controller, which indirectly indicates that the proposed control strategy has less current fluctuations. The maximum current difference of the proposed strategy is 68.93A, and the fuzzy control is 95.89A. This shows that the improved Savitzky-Golay filter can effectively smooth current fluctuations and delay battery degradation.

A. Battery Cycle Life Estimation Model

Experts define the cycle life of a battery as the number of discharge-charge cycles; After the battery capacity drops to 80%, it will be artificially terminated for battery service life. The battery's DOD, temperature and current rate have a great influence on the battery cycle life.

The basic motivation for using HESS is to extend the service life of the battery under frequent charging and discharging operations. So far, there have been a lot of prediction models lithium-ion battery capacity fade [31] [32] [8] [24] , these models are developed by the battery mechanism different directions, such as parasitic side reactions [24] and [25] increased resistance ,which will lead to the degradation of the battery. However, these models require extensive and sufficient experiments to obtain sufficient experimental data to study the aging process of the battery system and verify the mechanism of capacity degradation. Besides that, the calculation and calibration processes of these two prediction models are complicated, which makes it difficult to implement in actual electric vehicles. Wang and Liu proposed a semi-empirical life model that considers the influence of the four parameters of charge depth, temperature and discharge rate [31], as shown in Eq. (22). This model can clearly describe the factors that affect battery service life.

$$Q_{loss} = Ae^{-\left(\frac{E_a + B * C_{\_Rate}}{R * T_{bat}}\right)} (A_h)^\tau \qquad (22)$$

In the above equation, $Q_{loss}$ represents the degradation of battery capacity, $R$ is the ideal gas constant, $E_a$ is the activation energy from Arrhenius law (J / mol), $C_{Rate}$ is the absolute value of the battery charge and discharge current rate, $A_h$ is the Ah throughput, $A$ is the pre-exponential factor, $\tau$ is the second undetermined coefficient (Let the slope of the linear model be equal to 0.824.), $B$ is the undetermined third coefficient, and $T_{bat}$ is the Kelvin temperature of the battery.

Linearized by taking the natural logarithm on both sides of Equation (22), the result is

$$In(Q_{loss}) = InA - \frac{E_a + B * C_{Rate}}{R * T_{bat}} + \tau In(A_h) \qquad (23)$$

Numerous studies indicate that, with the increase in the current rate, the activation energy in the battery will be reduced. Therefore, set $E = E_a + B * C_{Rate}$ to represent the inverse relationship between activation energy and current rate. The parameters $B$ and $E_a$ are calibrated by the least square method, the calibration model is as follows.

$$E = 31500 - 370.3 * C_{Rate} \qquad (24)$$

According to the research in literature [24], the inverse relationship between activation energy and current rate box is shown in equation (22). In order to extend the cycle life of the battery model to estimate any case lower than 10C (generally lower than the electric vehicle battery magnification 10C), we have by 1 / 2C, 2C, 6C, $InA$ 10C rate values using curve fitting derived the relationship between the value and lnA magnification. The curve fitting result of $InA$ is shown in Figure 11. The fitting formula (25) is as follows.

$$InA = a * e^{-b * C_{Rate}} + c \qquad (25)$$

The data is fitted based on formula (6), and the least square method is used for calibration. Obtain the following relevant parameters, and the fitted graph is shown in Fig.11.

$$a = 1.251; b = 0.2539; c = 9.21 \qquad (25)$$

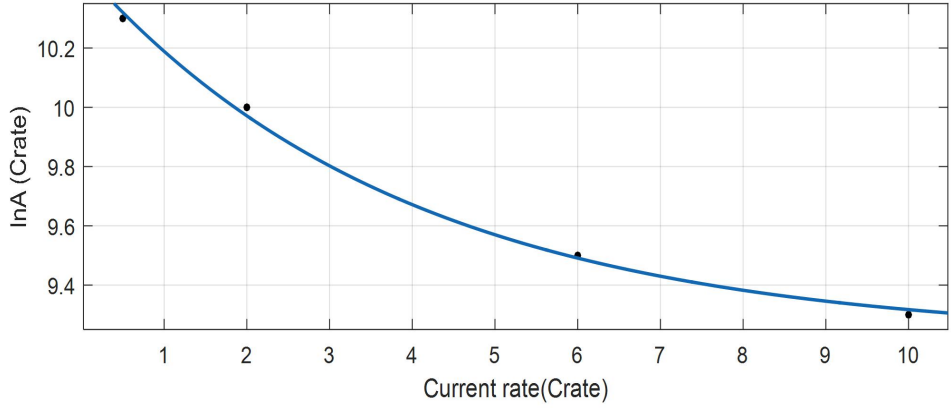

**Fig.11.** Fitting parameters of $lnA(C_{Rate})$ values under different current rates.

Assuming that the battery rate at time [p,p+1] is $C_{rate}(p)$, then the battery capacity loss at the $C_{rate}(p)$, can be estimated as

$$Q_{loss}(p) = Ae^{-\left(\frac{E_a + B * C_{Rate}(p)}{R * T_{bat}}\right)}(A_h(p))^\tau \qquad (26)$$

Therefore, the total capacity loss $Q_{loss}$ of the battery in the cycle period is the sum of the capacity loss under the conditions of different current rates.

$$Q_{loss} = \sum_{0}^{T} Q_{loss}(p) \qquad (27)$$

Under UDDS cycle conditions, $T$ is the time required for the test conditions. The battery life predicted by the model is shown in Fig.12 (based on 80% depth of discharge).

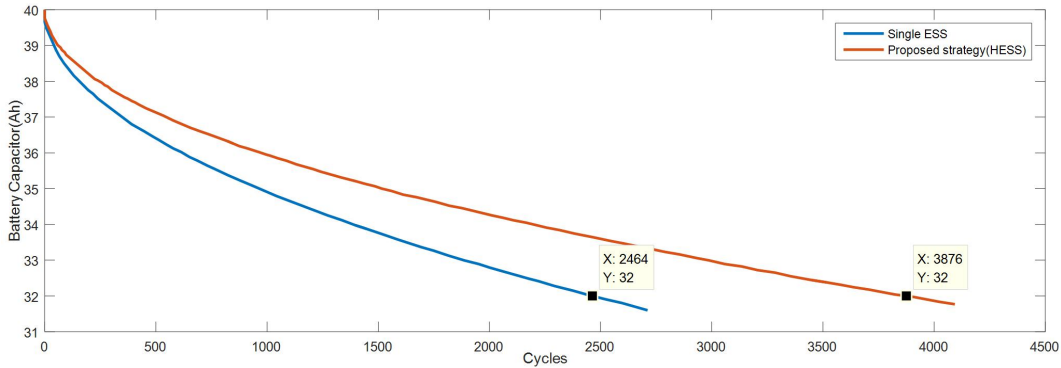

**Fig.12.** Estimate battery service life under UDDS.

As can be seen from Fig.12, HESS proposed strategy as compared to using only battery ESS, battery cycle life has been prolonged 57.31%.

B. The impact of different supercapacitor monomers on the system.

Supercapacitors have developed rapidly in recent years. Now researchers hope to develop supercapacitors with higher energy density that can better match with lithium batteries. This article uses a series of supercapacitor monomers to compose different HESS for simulation (the total amount of charges remains unchanged supercapacitor bank.), and estimate the impact of supercapacitor improvement on HESS. In Table V, A, B, C, and D use supercapacitor monomers on the market. E is the target monomer being studied by researchers. The calculation formula is as follows:

$$N_2 = \frac{C_1 * U_1 * N_1}{C_2 * U_2} \quad (28)$$

Among them, $C_1$ is the nominal cell capacity of capacitor A, $U_1$ is the nominal cell voltage of capacitor A, and $N_1$ is the number of cells of capacitor A; $C_2$ is the nominal cell capacity of capacitor K, and $U_2$ is the nominal cell voltage of capacitor K. (K represents any of the supercapacitors B, C, D, E.)

TABLE V
NUMBER OF SUPERCAPACITORS IN HESS

| serial number | Cell Voltage(V) | Cell Capacity(A) | Number of battery cells |
|---|---|---|---|
| A | 2.7 | 3000 | 95 |
| B | 2.7 | 3400 | 84 |
| C | 2.85 | 3400 | 80 |
| D | 3.0 | 3400 | 76 |
| E | 3.2 | 3800 | 64 |

The simulation results under the test cycle are as follows (Based on Advisor simulation):

TABLE VI
HESS SIMULATION RESULTS

|   | $SOC_{bat\_final}$ | $SOC_{sc\_final}$ | $Q_{bat\_loss}$ | $Q_{sc\_loss}$ | $Q_{total\_loss}$ | Percentage |
|---|---|---|---|---|---|---|
| A | 0.8949 | 0.5317 | 983.91 | 27.80 | 1011.71 | 0 |
| B | 0.8953 | 0.5300 | 975.54 | 27.77 | 1003.31 | 0.83% |
| C | 0.8955 | 0.5744 | 970.21 | 27.56 | 997.77 | 1.38% |
| D | 0.8959 | 0.6054 | 963.65 | 27.37 | 991.02 | 2.05% |
| E | 0.8969 | 0.7037 | 945.83 | 27.18 | 973.01 | 3.83% |

Can be seen from the Table VI, when the monomer of type A and B, two different capacity of supercapacitors monomer, large capacity of monomer can effectively reduce the energy loss and reduce the amount of monomer HESS(drops about 11.58%.), this means a reduction in the mass and volume of supercapacitors (general ultracapacitors monomer were similar or the same size and quality). B, C and D represent the voltage change of the monomer and the capacity of the monomer remains unchanged. With the increase of the voltage of the supercapacitor monomer, the energy loss of HESS was significantly reduced, and the number of monomer was also significantly reduced. Compared with capacitor A, supercapacitor D is approximately 20.00% less in mass and volume. Monomer E is the next target parameter pursued by the researchers, and it can be seen from the table VI that it reduces the number of ultracapacitors in HESS and simultaneously reduces the energy loss of HESS, which will make supercapacitor's mass and volume reduced about 31.58%. It is not difficult to see from these data that with the development of ultracapacitor, it can improve its advantage in HESS and further reduce HESS's energy loss.

## Conclusion

This paper proposes an intelligent energy management strategy based on fuzzy logic control-improved Savitzky-Golay filter to control the demand power distribution of electric vehicles in real-time, aiming to improve the usable cycle life of the battery by optimizing power distribution. By using model I and model II for selective fitting, the peak power of the battery can be effectively reduced, and the current fluctuation of the battery can be reduced. The improvement of these performance indicators helps to slow down the degradation of the battery and improve its service life.

In order to evaluate the performance of the proposed control strategy, MATLAB was simulated based on the test drive cycle. The simulation results show that the method can effectively improve various performance indicators of the battery. For example, in terms of the maximum temperature of the battery, the proposed control strategy is better than single ESS dropped by 8.47%. Besides, the control strategy proposed in HESS increases the service life of the battery by 57.31% compared to the single ESS, which helps to reduce the operating cost of electric vehicles. Through the exploration of the next-generation super capacitor, it is predicted that it will bring about a 31.58% reduction in volume and mass. Under the proposed strategy, it can be roughly seen that supercapacitors can bring better performance in the future.